\documentclass[aps,preprint,showpacs,superscriptaddress,groupedaddress]{revtex4-1}

\usepackage{color}

\usepackage[utf8]{inputenc}
\usepackage[french,english]{babel}
\usepackage[T1]{fontenc}
\usepackage[version=3]{mhchem} 
\usepackage{graphicx}
\usepackage{dcolumn}
\usepackage{bm}
\usepackage{xcolor}
\usepackage{caption}

\usepackage{amssymb}
\usepackage{amsmath}
\usepackage{epsfig}
\usepackage[colorlinks=false, pdfborder={0 0 0}]{hyperref}
\usepackage{cleveref}
\usepackage{siunitx}
\usepackage{tikz}
\usepackage{pgfplots}
\crefname{figure}{Figure}{Figures}
\crefname{equation}{Equation}{Equations}
\crefname{table}{Table}{Tables}

\begin{document}

%%%%%%%%%%%%%%%%%% title page information %%%%%%%%%%%%%%%%%%
\title{Remote electrical arc suppression by laser filamentation}

\author{Elise Schubert$^1$} 
\author{Denis Mongin$^1$} 
\author{J\'er\^ome Kasparian$^{2}$} \email{jerome.kasparian@unige.ch} %% email address is required
\author{Jean-Pierre Wolf$^1$}

\affiliation{$^1$GAP-Biophotonics, Universit\'e de Gen\`eve, Chemin de Pinchat 22, 1211 Gen\`eve 4, Switzerland}
\affiliation{$^2$GAP-Biophotonics, Universit\'e de Gen\`eve, Chemin de Pinchat 22, 1211 Gen\`eve 4, Switzerland}

\date{\today}

\begin{abstract}
We investigate the interaction of narrow plasma channels formed in the filamentation of ultrashort laser pulses, with a DC high voltage. The laser filaments prevent electrical arcs by triggering corona that neutralize the high-voltage electrodes. This phenomenon, that relies on the electric field modulation and free electron release around the filament, opens new prospects to lightning and over-voltage mitigation.  
\end{abstract}

\maketitle

%%%%%%%%%%%%%%%%%%%%%%%%%%  body  %%%%%%%%%%%%%%%%%%%%%%%%%%
\section{Introduction}

Since the pioneering work of Benjamin Franklin~\cite{Krider2004}, both the mechanisms of lightning and lightning protection have entered in the area of science. 
Lightning occurs after vertical winds, with speeds up to 20~m/s, separate charges within thunderclouds. The resulting static electric fields reach 10 -- 15~kV/m at ground level, and up to 50~kV/m some hundreds of meters above ground. These high fields initiate corona discharges, that under sufficient electric field and charge supply develop into streamers, and then ionized plasma channels, or leaders. The latter give rise to a lightning strike if they connect regions of clouds and/or of the ground with opposite charges~\cite{Gary2004}.

Lightning protection of buildings or other facilities is mainly focused on capturing the lightning strikes and safely conducting their energy to the earth, via a grounding conductor. Alternatively, it has been proposed to use lasers to divert lightning to harmless places by guiding it using the ionized path generated by laser pulses~\cite{MikiAS1993,ZhaoDWE1995,UchidSYMYYKT1999,ComtoCDGJJKFMMPRVCMPBG2000,Bazelyan2000,LaCCDGJJKMMPRVPCM2000,RamboSD2001,Apollonov2002,RodriSWWFAMKRKKSYW2002,KaspaRMYSWBFAMSWW2003,KaspaAAMMPRSSYMSWW2008a}. Laser filaments have also been used as a probe to investigate the dynamics of high-voltage discharges \cite{SugiyFMYZHN2009,Sugiyama2010} and leaders~\cite{EtoZOMF2012,Schmitt-Sody2015,arXivWang2015}.

Laser filamentation~\cite{ChinHLLTABKKS2005,CouaiM2007,BergeSNKW2007} is a self-guided propagation regime of high power, ultrashort pulses. It relies on an interplay between the self-focusing Kerr effect and a defocusing effect caused by ionization of the medium, as well as higher-order Kerr effects \cite{BejotKHLVHFLW2010a,B'ejHLKWF2011a,BejCHLKWF2013,Weerawarne2015}. The resulting dynamic balance  clamps the filament intensity to ~5~$\times$~10$^{13}$~W/cm$^3$ \cite{KaspaSC2000,BeckeAVOBC2001} over a typical diameter of ~100~$\mu$m, producing a continuously ionized plasma channel. 

Most of the work on laser triggering of high-voltage discharges to date uses high-voltage pulses synchronized with the laser, unlike the quasi-DC field typical of thunderclouds. Furthermore, triggering lightning ultimately implies collecting the  discharge by means of a classical lightning rod, with an adequate design and/or location to avoid indirect effects of lightning, i.e., electromagnetic perturbation, in the vicinity of the impact location. 

In contrast, preventing breakdown at all by remotely unloading thunderclouds may offer a more efficient protection. In that regard, grounded wires of 100~$\mu$m or less have been shown to inhibit electrical arcs by initiating a streamer-free corona (or ultra-corona)~\cite{Rizk2010}. Visible with naked eye as a glow discharge, ultra-corona disperses a space charge that stops sparking.
Here, we show that laser filaments can induce a similar process and neutralize DC high-voltage electrodes while preventing electrical arcs, even remotely. 

\section{Experimental setup}

\begin{figure}[ht]
\begin{center}
\includegraphics[width=12cm, keepaspectratio]{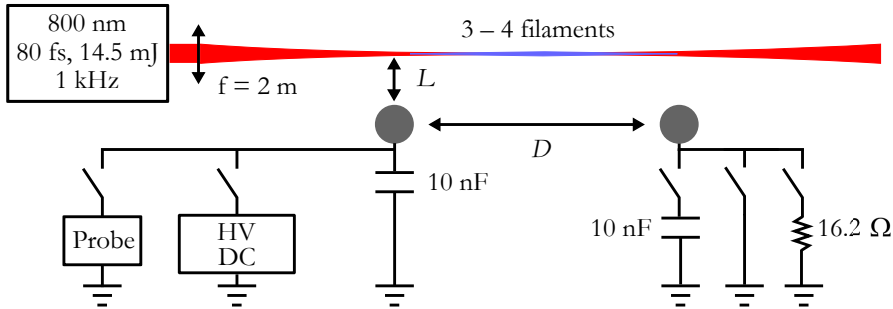}
\end{center}
\caption{Experimental setup. A Ti:Sapphire laser focused with an $f=2$~m lens creates 3--4 $\sim20$~cm long filaments. The electrode with a capacitor is charged to 14~kV and the generator is then disconnected. After the laser has been shined, the residual voltage on the electrode is measured via a HV-probe. The distance between the electrodes and the distance from the electrodes to the filament can be changed.} \label{fig:Setup}
\end{figure}

\noindent The experimental setup (\cref{fig:Setup}) relied on a DC high-voltage generator (100~kV, 200~$\mu$A maximum current, or 14~kV, 7~mA with a 10~nF capacitor attached to store charge and allow the delivery of stronger currents), connected to spherical electrodes (diameter $1.2$~cm) that could be placed arbitrarily on either side of the laser beam, at various axial and longitudinal positions. Unless otherwise specified, one electrode was grounded, while the other one was set to the potential of the high-voltage generator.
Also, the generator could be disconnected from the capacitor and the electrode, so as to investigate the flow of a fixed charge $Q$ between the electrodes and the associated unloading of the capacitor.

A Ti:Sapphire chirped pulse amplification system delivered 80~fs pulses of 14.5~mJ centered at 800 nm at a 1~kHz repetition rate, unless otherwise specified. This beam of initial diameter 4~cm was slightly focused ($f$~=~2~m), generating 3 to 4 self-guided filaments. Due to the external focusing of the beam, the filament length was about 20~cm in our configuration. As detailed in the next section, the beam was sent in various configurations relative to the electrodes, from a close vicinity of 2~mm (the minimum distance to avoid the edge of the beam to ablate matter from the electrodes) to 30~cm distance.

The current flowing through the ground electrode was measured by monitoring the voltage on a 16.2~$\Omega$ resistor connecting  the latter to the earthing. The signal was recorded on a 100~MHz bandwidth oscilloscope, synchronized with the laser pulses by a photodiode detecting the scattering of the beam on the beam dump, $2.2$~m downstream of the interaction region. 
The discharge of the capacitor attached to the high-voltage electrode was characterized using a high-voltage probe (Fluke 80K-40 HV probe).

The experiments have been performed at room temperature ($T = 20^{\circ}$C), at a relative humidity around 30\%. Correspondingly, the background resistivity of the air is about 3$\times$10$^{14}$~$\Omega\cdot$m~\cite{Pawar2009}.

\begin{figure}[t]
\centering
\includegraphics[width=12cm, keepaspectratio]{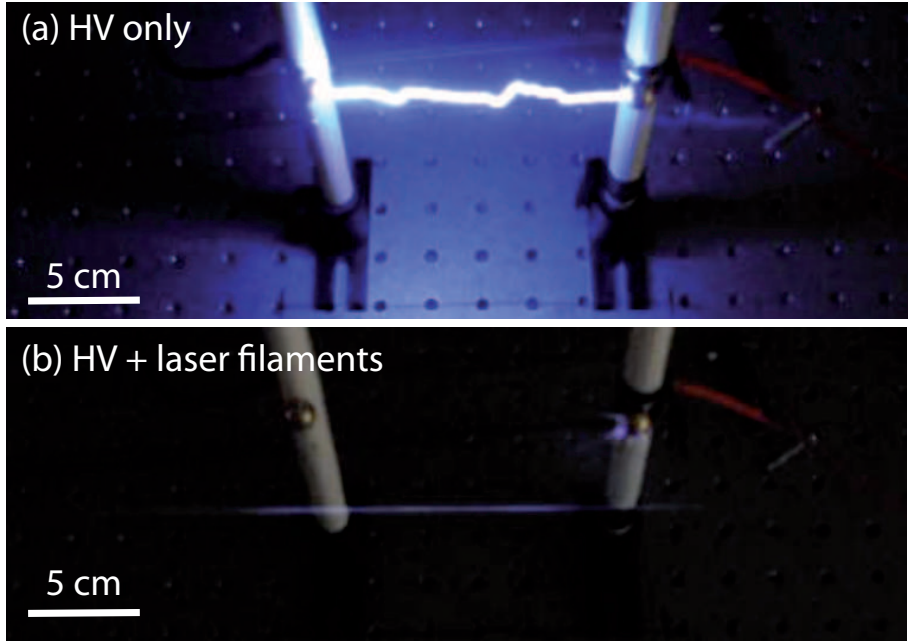}
\caption{Electrical arc suppression by laser-induced neutralization under 100~kV %(See {\textcolor{blue}{Visualization 1}}). 
(a) Electrical arc without laser (b) Arc inhibition when the laser is turned on. 
} \label{fig:photos}
\end{figure}

\section{Results}
\subsection{Electrical arc suppression}
Figure~\ref{fig:photos} and the Supplementary movie %({\textcolor{blue}{Visualization~1}) 
clearly illustrate electrical arc suppression by laser filaments. Applying 100~kV (\cref{fig:photos}(a)) in a 12~cm gap between the electrodes, a value close to the threshold for spark discharges, results in typically one electrical arc per second. Switching on the laser in the vicinity of the electrodes immediately suppresses the sparking (\cref{fig:photos}(a)). Instead, a glow discharge connects the electrodes to the filament, even if the latter is several centimeters away from the electrodes. 
Furthermore, the blueish glow of the filament, that is typical of the laser-generated plasma~\cite{CouaiM2007,BergeSNKW2007}, is much brighter in presence of the high-voltage than without it evidencing increased avalanche ionization in the filament under the electric field. 

\begin{figure}[t]
\centering
\includegraphics[width=12cm, keepaspectratio]{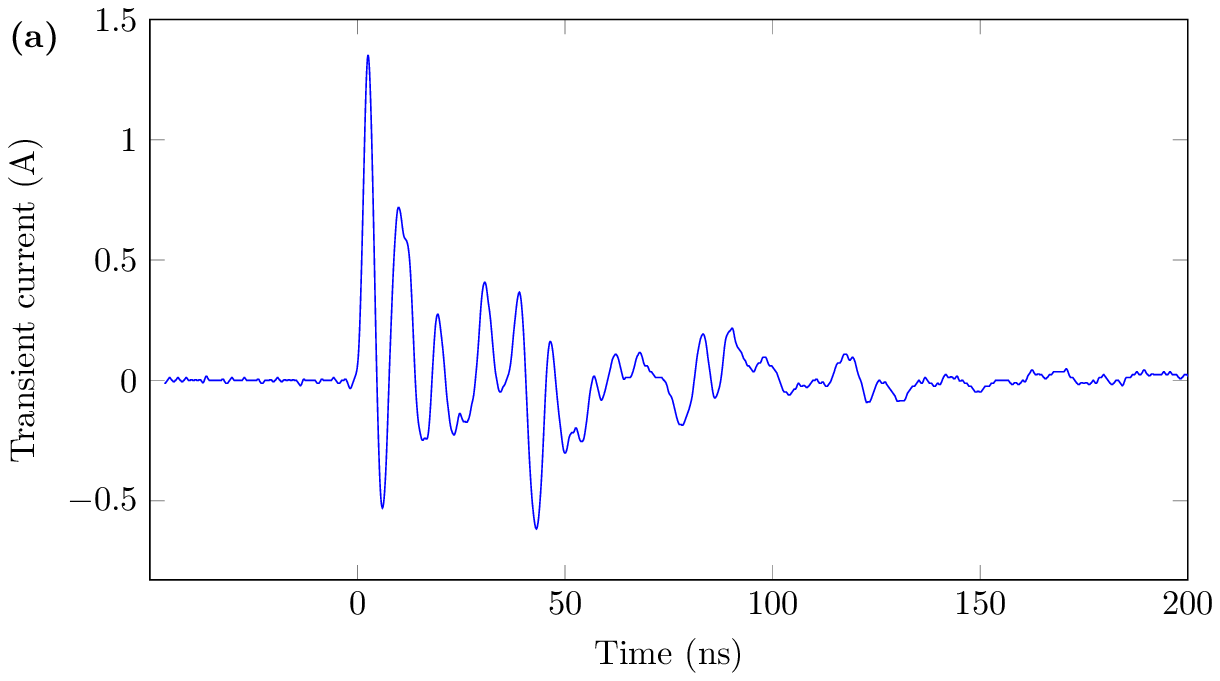}
\includegraphics[width=12cm, keepaspectratio]{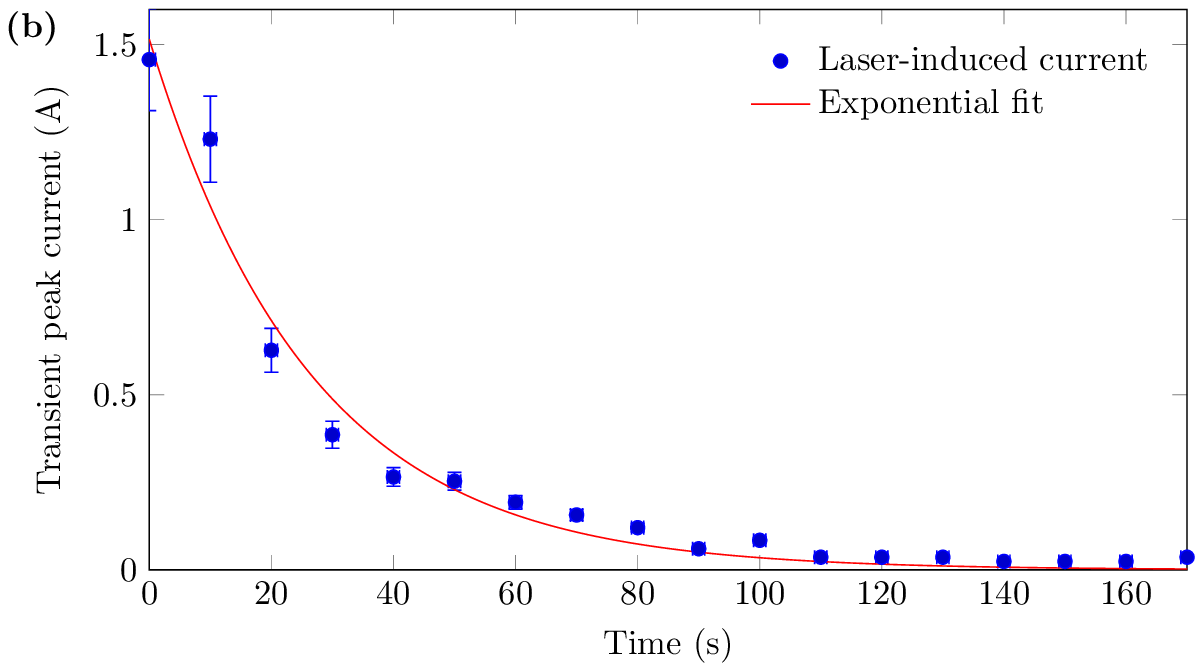}
\caption{Laser unloading of a 10~nF capacitor at +14~kV. The electrodes were placed at a distance $D$~$= 16\pm0.2$~cm apart, at a lateral distance $L$ of $2$~mm from the laser beam. (a)~Laser-induced current. Time 0 corresponds to the laser pulse. (b) Decay of the magnitude of the peak current, corresponding to the capacitor discharge.} \label{fig:decharge}
\end{figure}

\subsection{Spark-free neutralization}
To better understand the electrical arc suppression and the associated glow discharge, we investigated the current flow induced by the laser filaments between the two electrodes with a 10~nF capacitor on the HV side, loaded under +14~kV.  
Without the laser, no measurable current flows between the electrodes, so that the capacitor keeps its load without measurable leak for over 20~hours. In contrast, each laser pulse results in a current burst (Figure~\ref{fig:decharge}(a)). The time delay between the laser pulse and the initiation of the corresponding current burst is shorter than the 5~ns time resolution of our detection system. It is therefore faster than the time required to establish a negative corona discharge, that lies in the 100~ns range~\cite{SugiyFMYZHN2009}.

These bursts progressively unload the capacitor (\cref{fig:decharge}(b)), with a decay time $\tau = 25$~s, corresponding to 25000 laser pulses. The laser-induced current is still observed when the laser is shifted laterally up to $L$~=~30~cm away from the electrodes. 

\begin{figure}[t]
\centering
\includegraphics[width=12cm, keepaspectratio]{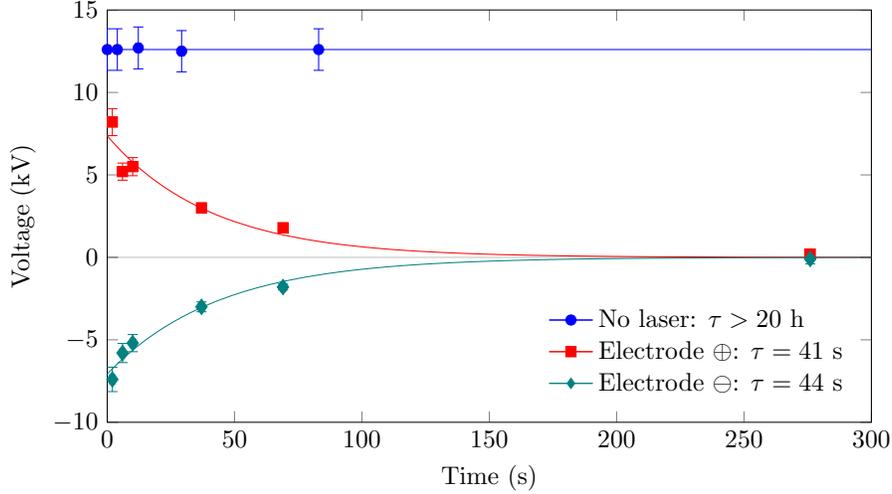} 
\caption{Simultaneous discharge of capacitors loaded under $\pm14$~kV, with a 16~cm gap between the electrodes, placed at $L=2$~mm from the laser. The lines display exponential fits, $\tau$ being the decay time.\label{fig:Symetrique}}
\end{figure}

We characterized the flow of charges from one electrode to the other by attaching an initially empty capacitor $C_2$ to the ground electrode (see \cref{fig:Setup}). Under laser influence, the high-voltage capacitor $C_1$ progressively discharges. About one third of the charge from $C_1$ indeed reaches $C_2$. As a result the charge of the latter reaches a maximum after 81~s, before decaying again. 

In contrast, when both 10~nF capacitors are simultaneously pre-loaded under $+14$ and $-14$~kV, respectively, in a floating configuration, they unload symmetrically (\cref{fig:Symetrique}), independently from the electrode polarity. 

\begin{figure}[t]
\centering
\includegraphics[width=16cm, keepaspectratio]{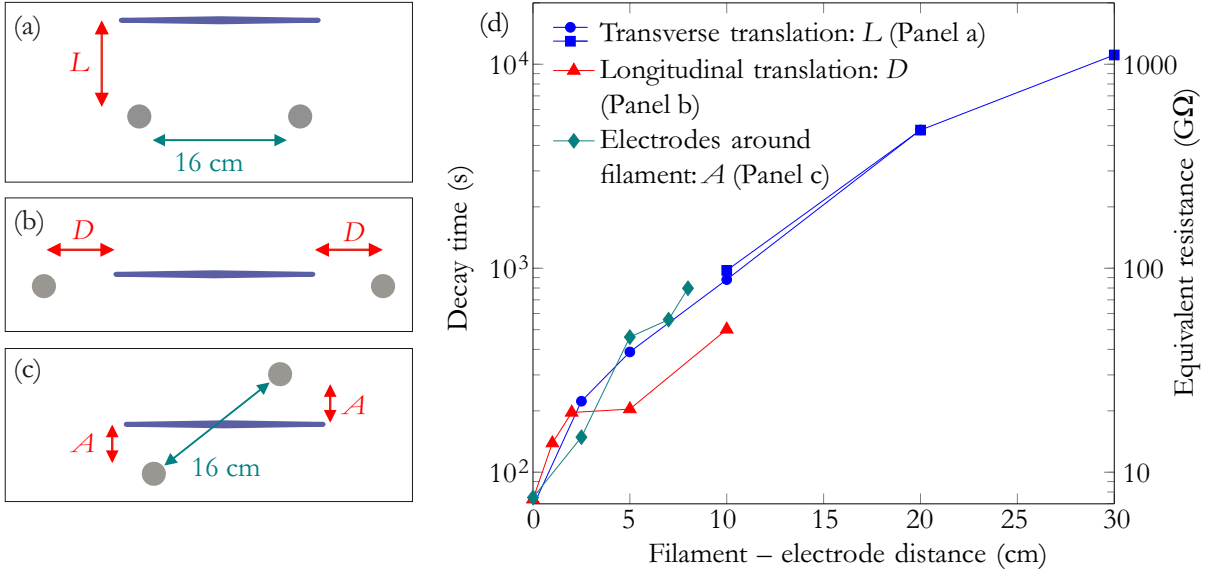}
\caption{Effect of geometry on laser-induced neutralization. 
(a-c) Geometrical configurations: (a) transverse and (b) longitudinal distances, and (c) filament crossing the electrode gap. The gray circles represent the electrodes, and the blue line is the filament.
(d) Decay time of the charge as a function of the distance between the electrodes and the filaments. 
} \label{fig:distance}
\end{figure}

\subsection{Geometrical considerations}

Associated with electrical arc inhibition, the laser filaments can thus induce current and consequently neutralize the high-voltage even remotely. More precisely, we investigated three geometrical configurations. In the first one, the laser beam was moved away laterally from the electrodes and parallel to the gap (\cref{fig:distance}(a)). In the second one, the electrodes were taken apart from one another, beyond the filament ends: the filament-distances therefore increased longitudinally  (\cref{fig:distance}(b)). Finally, we rotated the electrodes on a 16~cm axis centered around the filament. Tilting the axis varied the filament--electrode distance  (\cref{fig:distance}(c)).

In these three configurations, the decay times (\cref{fig:distance}(d)) are similar for identical distances between the electrodes and the filament.
Therefore, the distance between the filament and the electrodes, rather than the geometrical configuration, governs the laser-induced neutralization. In particular, setting the filament parallel or perpendicular to the electrode gap results in similar laser-induced neutralization times. 

Furthermore, the decay time of the charge $Q_0=C\cdot V=1.4\times10^{-4}$~C increases with the filament--electrode distance, from 68~s at 0.2~cm, to 3.1~h at 30~cm. This increase is linear up to a typical filament--electrode distance of 10~cm. Beyond this distance, the decay time tends to grow quadratically with increasing filament--electrode distances, while the decay of the charge progressively deviates from exponential, suggesting the occurrence of a new regime at long distances. However, even for a filament--electrode distance of 30~cm equal to twice the gap between the electrodes, the laser accelerates the unloading of the setup by one order of magnitude, illustrating the long-range effect of the laser filaments.

The exponential decay of the charge, together with its linear dependence with distance at least over the first 10~cm, are consistent with the simple picture of the discharge of an $RC$ circuit. Within this rough, quasi-stationnary description, the observed decay times correspond to effective resistances of 6.8~G$\Omega$ to 1~T$\Omega$, since $C$ = 10~nF, as displayed on the right vertical axis of \cref{fig:distance}. 
Note that, to keep reasonable experimental times, the measurement at 30~cm has been performed with a capacitor reduced to 200~pF, and then renormalized to $C$~=~10~nF. We checked for 10 and 20~cm distances that this procedure yields consistent results (blue squares in \cref{fig:distance}(d)).  

Still, for a given filament--electrode distance, the charge decays slightly faster when the beam passes close to the electrodes and the filament bridges part of the electrode gap (\cref{fig:distance}(b)), as compared to the other configurations.  This may be due to the longitudinal plasma density along the filament, that rises faster at the beginning of the filament than it falls down at its end~\cite{CouaiM2007,BergeSNKW2007}. As a consequence, the ends of the ionized region cannot be precisely defined.

\begin{figure}[t]
\centering
\includegraphics[width=12cm, keepaspectratio]{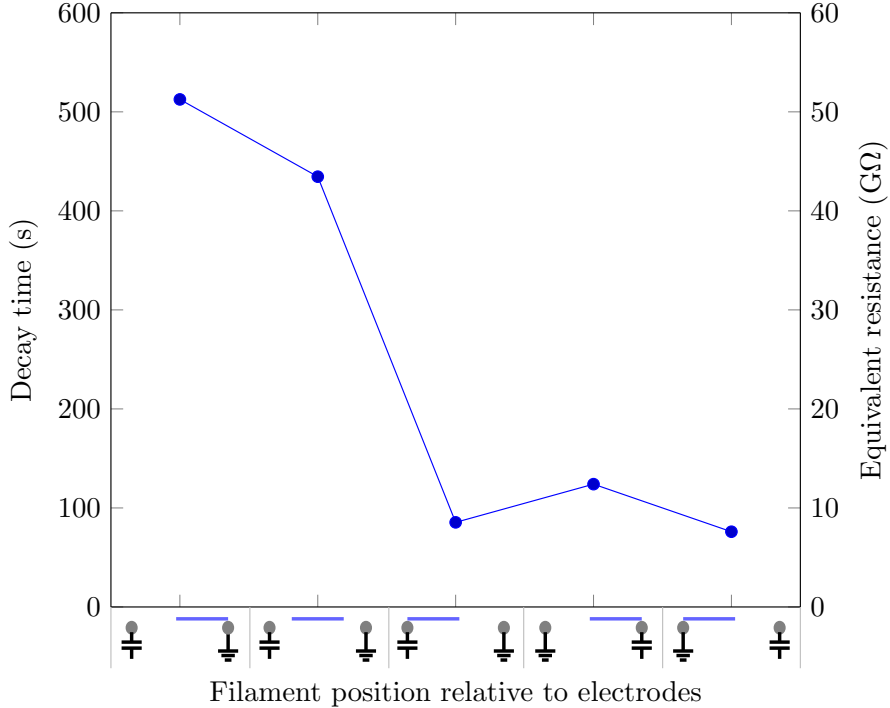}
\caption{Charge decay time as a function of the relative position of the filament and the electrodes, located 2~mm away from the laser beam. On the symbols of the $x$-axis, the laser filament (in blue) propagates from left to right. The capacitor is charged at $+14$~kV and the gap between the electrodes is 40~cm.} \label{fig:Position}
\end{figure}

As a result of the longitudinal electron density profile asymmetry in the filament, the charge decays 6.5 times faster when the high-voltage electrode is located 20~cm before the filament than at the same distance beyond it (\cref{fig:Position}). This strongly asymmetric behavior illustrates the key role played by the high density of free charges released in the laser filament. In contrast, the neutralization is insensitive to the position of the grounded electrode, as well as to the longitudinal position of the electrodes within the filament. 

Finally, we note that the unloading of a negative electrode facing a ground electrode behaves similar to that of a positive electrode \emph{vs.} ground.

\section{Discussion}

As the filament offers a long conductor of 100~$\mu$m diameter similar to ultra-corona generating devices used to prevent sparking~\cite{Rizk2010}, it can be expected that its effect relies on a similar mechanism. 
Offering a sharp conducting surface, ultra-corona devices strongly deform and enhance the electric field locally.
In the case of anode corona, the electrons are accelerated towards the electrode, leaving a space charge of positive ions behind them. Furthermore, they generate secondary charges by avalanche ionization in the air, as well as impact ionization on the electrode surface~\cite{Uhlig1956,Rizk2010,Fizk2014}. The local positive charge left behind by the electrons locally creates a strong electric field that in turn can field-ionize additional air molecules further apart.
The resulting charge flow is visible as a glow discharge and audible by naked ear. Furthermore the partial screening of the electric field drastically increases in the breakdown voltage, similar to the charge neutralization and the electrical arc suppression by the filaments.% (See {\textcolor{blue}{Visualization~1}).

However, filaments differ from a conducting wire in several ways. 
They generate a long, connected 100~$\mu$m diameter plasma with an electron density of at least $\rho_{\textrm{fil}}$~=~10$^{21}$~m$^{-3}$  and a lifetime in the $\mu$s range~\cite{CouaiM2007,BergeSNKW2007}, seven orders beyond typical values in a corona, $\rho_{\textrm{c}}~=~$10$^{14}$~m$^{-3}$~\cite{Chen2002}. These electrons are free to spread around the laser beam, not being confined to the initial conducting volume. 
As a consequence, much more space charges can be generated over much larger volumes. Furthermore, during the electron lifetime, their displacement can strongly affect the electric field, and induce significant currents.

A rough estimate of this particular aspect can be obtained by considering the electron mobility~\cite{ZhaoDWE1995}:
\begin{equation}
\mu_e (m^2/V\cdot s) = -\frac{N_0}{3N}\left(\frac{5\times10^5+E_0}{1.9\times10^4+26.7\times E_0}\right)^{0.6}
\label{eq:mobility}
\end{equation}
where $E_0=E N_0 / N$, $E$ is the electric field, and $N$ and $N_0$ are the molecule densities in the considered condition and in normal conditions, respectively. 

As a result, if we neglect ionic mobility, the resistance of the plasma over a pathway $L$ expresses as:

\begin{equation}
R = \frac{1}{e \mu_e S}\int_0^L{\frac{\textrm{d}\vec{r}}{\rho_{\textrm{e}}(\vec{r})}}
\end{equation}
where $S$ is the cross-section of the conducting area and $\rho_e$ the free electron density. 

According to \cref{eq:mobility}, $\mu_e$~=~0.2 m$^2$/V$\cdot$s for an electric field of 50~kV/m, incuding  electron drift speeds close to 10$^4$ m/s, corresponding to 10~cm within 1~$\mu$s, a typical plasma lifetime in the electric field \cite{TzortPFM2000}. We therefore expect a monotonic decay of the electron density for increasing distances $r$ around the filament. Assuming an exponential decay, the electron density between the filament and the electrode, at a distance $r$ from the filament, is modeled by   
 ${\rho_e(r) =\rho_{\textrm{fil}} \exp\left(-r \ln \rho / L \right)}$, with $\rho=\rho_{\textrm{fil}}/\rho_{\textrm{c}}$. 
As a consequence,
\begin{eqnarray}
R = \frac{L}{e \mu_e S \rho_{\textrm{fil}}}\frac{\rho-1}{\ln \rho} \approx \frac{L}{e \mu_e S \rho_{\textrm{fil}}}\frac{\rho}{\ln \rho}
\end{eqnarray}
which amounts to 20~M$\Omega$ for a cross-section $S$~=~1~cm$^2$ typically corresponding to a column of air with a cross-section comparable to that of the electrodes, and a length $L$~=~10~cm. 

However, the $\mu$s plasma lifetime~\cite{TzortPFM2000} is much shorter than the ms time between two laser pulses at a repetition rate of 1000~Hz. This 1/1000 duty cycle will result in a 1000-fold higher effective resistance. 
Furthermore, the current measured in our experiment flows from the HV electrode to ground electrode, via the filament and the two gaps between the electrodes and the filament. Consequently, our crude estimate yields a value of 40~G$\Omega$ for the whole path when the filament is 10~cm away from the electrodes. This value matches well the experimental one (See \cref{fig:distance}). 
We note that the resistance of the filament itself has been neglected, as it lies in the M$\Omega$/m range~\cite{RodriSWWFAMKRKKSYW2002}, well below the gap between the filament and the electrodes.

Although the above discussion disregards the plasma dynamics, including the plasma lifetime and the avalanche, and it does not consider the contribution of the heavy species both to the field deformation and to the current flow, this estimation illustrates the critical contribution of the high electron densities generated in the filament on the electrical arc suppression by ultra-corona generation. It also explains why the two ends of the filament behave differently (\cref{fig:Position}), related with the differences in their longitudinal electron density profiles.

Another specificity of the filament as compared to the wire-based ultra-corona generating devices is that the former is not grounded.
Although it may be surprising that grounding is not necessary to generate ultra-corona, it is reminiscent of the triggering of lightning by rockets pulling only a short section of conducting wire, without connection to the ground. The local exaltation of the electric field is sufficient to initiate the dischrage~\cite{RakovUR2005}.
Furthermore, filaments provide a large amount of charge available to neutralize the high-voltage, not requiring supply from the ground. 

More specifically, one can roughly estimate the amount of charge available for the neutralization, e.g., considering the conditions of \cref{fig:Symetrique}, where a 10~nF capacitor loaded under 10~kV is unloaded in typically 40~s. The 3--4 filaments of our setup neutralize $Q_0$~=~10$^{-4}$~C.
 In comparison, the charge produced in a 16~cm-long section of 100~$\mu$m diameter, with an electron density $\rho_{\textrm{fil}}~=~10^{21}$~m$^{-3}$ in a single filament amounts to 200~nC. Over 40~s, the setup will therefore have produced 8~mC. This value lies two orders of magnitude beyond the initial charge to neutralize. It may even be underestimated as the electron density generated by the filaments could be higher.

\section{Conclusion}

As a conclusion, we have observed suppression of electrical arcs by laser filaments. 
This effect is efficient even remotely, up to distances at least twice the gap between the electrodes. Filaments generated by ultrashort laser pulses provide a 100~$\mu$m thin conducting channel. Similar to metallic wires of comparable diameter producing ultra-corona \cite{Rizk2010}, they locally enhance the electric field, releasing high concentrations of space charge that allow current to flow and neutralize the electrodes, reducing the electrode potential and preventing sparking.

This new phenomenon opens the way to new approaches to lightning and over-voltage protection. As compared with a metallic thin wire, filaments generated in the air are generated by each laser pulse and therefore avoid any problem of mechanical resistance. 

\section*{Acknowledgments} 
We acknowledge financial support from the ERC advanced grant "Filatmo". We thank I.~Crasse, D. Eeltink, S. Hermelin, M.~Matthews, and M.~Moret for experimental assistance, and N.~Berti for numerical support.

%%%%%%%%%%%%%%%%%%%%%%% References %%%%%%%%%%%%%%%%%%%%%%%%%

%\end{thebibliography}

\newpage
\section*{Erratum}

In the discussion of~\cite{Schubert2015}, we erroneously converted the $10^4$~m/s drift speed of the electrons into a \SI{1}{\micro\s} transit time through the $L$~=~\SI{10}{\cm} gap between the filament and the electrodes. The correct transit time of \SI{10}{\micro\s} is well beyond the electron lifetime, preventing them to reach the electrode and ensure the conduction, at least in a streamer at a temperature close to ambient. As a consequence, their distribution cannot be modelled as an exponential decay between the filament and the electrodes, so that Eq. (3) is irrelevant and electrons cannot ensure the observed remote unloading under the experimental conditions of~\cite{Schubert2015}. 

Rather, the ions have a mobility of \SI{2e-4}{m^2/\V\s}~\cite{ZhaoDWE1995}, hence a drift  speed \SI{10}{m/s} under an electric field of \SI{50}{kV/m}. Therefore, they  transit in \SI{10}{ms}. After that time, their concentration is still \SI{2e19}{m^{-3}} according to the plasma model described in~\cite{Schubert2016}. This ion density is sufficient to ensure the observed conductivity of several tens of \si{G\ohm}~\cite{Schubert2015}, a conclusion consistent with recent results on electric probing of filaments~\cite{Polynkin2012}. However, in the case of large electric fields leading to a leader-like regime~\cite{FujiiMGZFON2008}, a contribution of the electrons along with the ions may still be  considered.

\end{document}